\documentclass{ws-procs9x6}         % default, citations in superscript
\begin{document}
\title{Using TDHF to study quasifission dynamics}
\author{A. S. Umar$^*$}
\address{Physics and Astronomy, Vanderbilt University,
Nashville, TN 37235, USA\\
$^*$E-mail: umar@compsci.cas.vanderbilt.edu}
\author{C. Simenel}
\address{Department of Nuclear Physics, The Australian National University,\\ Canberra ACT 2601, Australia}

\begin{abstract}
We show that the microscopic TDHF approach provides an important tool to
shed some light on the nuclear dynamics leading to the formation of superheavy elements.
In particular, we discuss studying quasifission dynamics and calculating ingredients for %$P_{\mathrm{CN}}$ 
compound nucleus formation probability calculations.
\end{abstract}


\bodymatter

\section{Introduction}\label{aba:sec1}
%The Hartree-Fock approximation
%and its time-dependent generalization, 
The time-dependent Hartree-Fock (TDHF)
theory has provided a possible means to study the diverse phenomena
observed in low energy nuclear physics~\cite{simenel2012}.
As a result of theoretical approximations (single Slater determinant), 
TDHF %does not
describes reaction channels in a common mean-field, usually corresponding to the dominant reaction channel.
For instance, it describes above-barrier fusion and average transfer dynamics~\cite{simenel2008}.
To obtain multiple reaction channels or widths of
observables one must go beyond TDHF~\cite{tohyama2001,lacroix2014,simenel2011}.
In connection with
superheavy element formation, the theory predicts
best the cross-section for a particular process which dominates the
reaction mechanism. This is certainly the case for studying capture
cross-sections and quasifission.

In recent years has it become numerically feasible to perform TDHF calculations on a
3D Cartesian grid without any symmetry restrictions
and with much more accurate numerical methods~\cite{umar1991a,maruhn2014}.
During the past several years, a novel approach based on TDHF called the density constrained
time-dependent Hartree-Fock (DC-TDHF) method was developed to compute
heavy-ion potentials~\cite{umar2006a,umar2014a} and excitation energies~\cite{umar2009a} directly from TDHF
time-evolution. This method was applied to calculate capture cross sections for
fusion reactions leading to superheavy element $Z=112$~\cite{umar2010a}.
Furtermore, within the last few years the TDHF approach has been utilized for studying the dynamics of
quasifission~\cite{wakhle2014,oberacker2014,umar2016,umar2015a,umar2015c,hammerton2015,sekizawa2016}.
The study of quasifission is showing a great promise to provide
insight based on very favorable comparisons with experimental data.

\section{Recent quasifission studies}

One of the major questions that is asked by the experimental superheavy element community is
why a $^{48}$Ca beam is so crucial in forming such systems and whether
one could produce new superheavy nuclei using projectiles different than
$^{48}$Ca and actinide targets. Our first work in this area focused on the
quasifission studies for the $^{40,48}$Ca+$^{238}$U system\,\cite{wakhle2014,oberacker2014},
showing that for neutron-rich $^{48}$Ca beams quasifission is substantially
reduced. While, the above work is also in the time span of this proposal, below
we discuss the more recent studies of the $^{48}$Ca+$^{249}$Bk and $^{50}$Ti+$^{249}$Bk systems\,\cite{umar2016}.
The reaction $^{48}$Ca +$^{249}$Bk creates superheavy isotopes of
element $117$ with cross-sections of 2--3 picobarns. However, the $^{50}$Ti +$^{249}$Bk
reaction so far has not produced any superheavy isotopes of element $119$, with
an upper cross-section limit of $50$~fb\,\cite{dullmann2013}.

We have calculated the microscopic DC-TDHF nucleus-nucleus potential barriers
for the $^{48}$Ca+$^{249}$Bk and $^{50}$Ti+$^{249}$Bk systems 
%The barriers are calculated 
for two extreme orientations
of the $^{249}$Bk nucleus (tip and side). %As expected, 
For the $^{48}$Ca+$^{249}$Bk system,
the tip orientation of $^{249}$Bk
results in a significantly lower barrier, $E_B$(tip)$=191.22$~MeV, %located at internuclear
%distance $R_B$(tip)$=15.04$~fm, 
as compared to the
side orientation, $E_B$(side)$=204.36$~MeV. % with $R_B$(side)$=12.47$~fm.
This reaction
has been studied at $E_\mathrm{c.m.}=$~204--218~MeV in  Dubna\,\cite{oganessian2013} and
$E_\mathrm{c.m.}=$~211--218~MeV with GSI-TASCA\,\cite{khuyagbaatar2014}. 
%We conclude that 
Thus, the highest experimental energy $E_\mathrm{c.m.}=218$~MeV is
above both barriers but the lowest experimental energy $E_\mathrm{c.m.}=204$~MeV
is slightly below the barrier for the side orientation of $^{249}$Bk.
Similarly, the corresponding potential barriers
for the $^{50}\mathrm{Ti}+^{249}\mathrm{Bk}$ system %the tip orientation of $^{249}$Bk
%results in a significantly lower barrier, 
are $E_B$(tip)$=211.2$~MeV %located at internuclear distance $R_B$(tip)$=14.48$~fm, as compared to the side orientation,
 and $E_B$(side)$=224.6$~MeV. % with $R_B$(side)$=12.96$~fm.
Experimentally, $E_\mathrm{c.m.}=233.2$~MeV
was used in the GSI-TASCA experiment\,\cite{khuyagbaatar2014}, well above both barriers.
%We observe that the chosen experimental energy is $22.0$~MeV above the barrier $E_B$(tip)
%and $8.6$~MeV above the barrier $E_B$(side).
\begin{figure}[!htb]
\begin{center}
\centering \includegraphics*[scale=0.26]{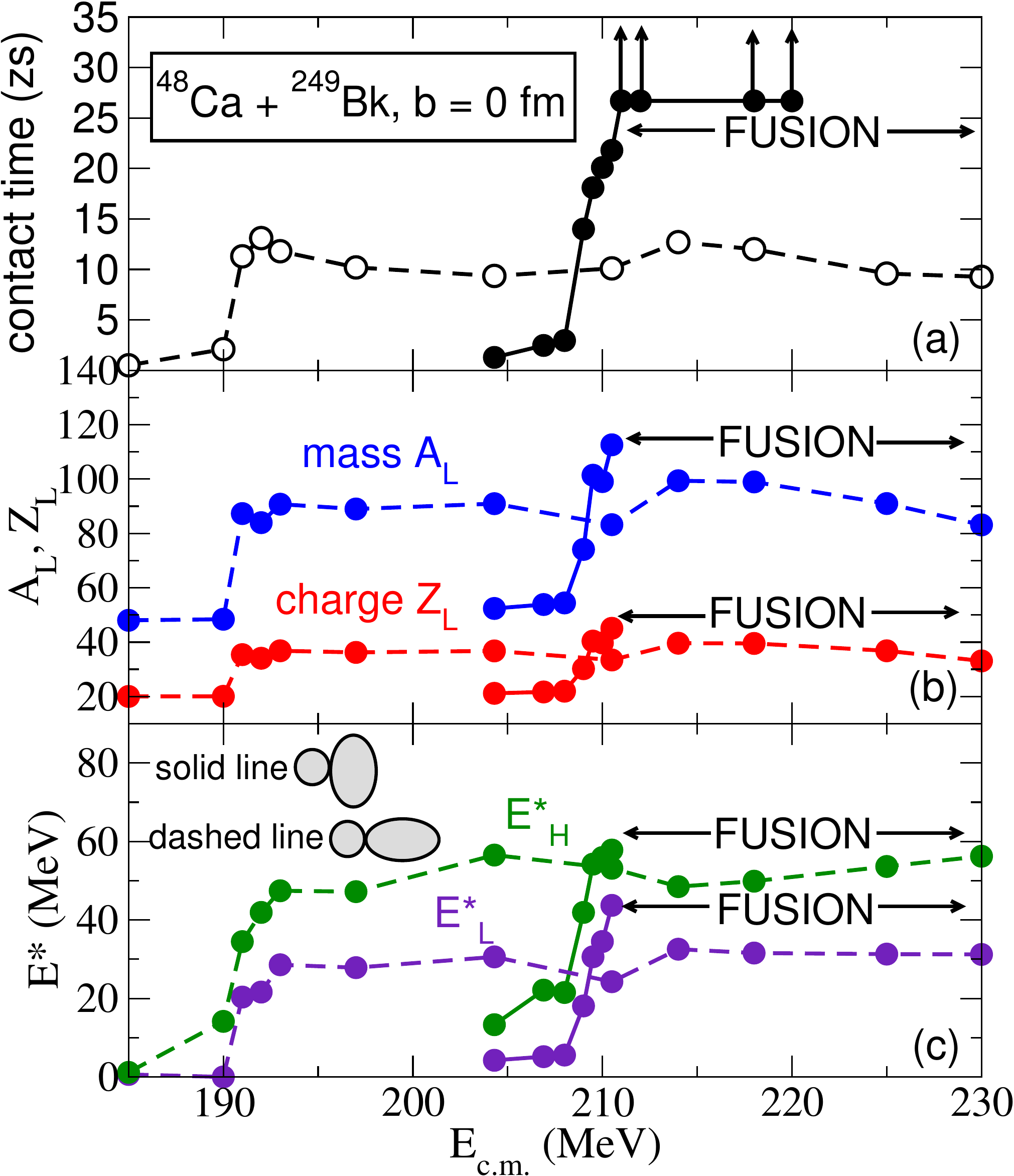}\hspace{0.1in}
\centering \includegraphics*[scale=0.26]{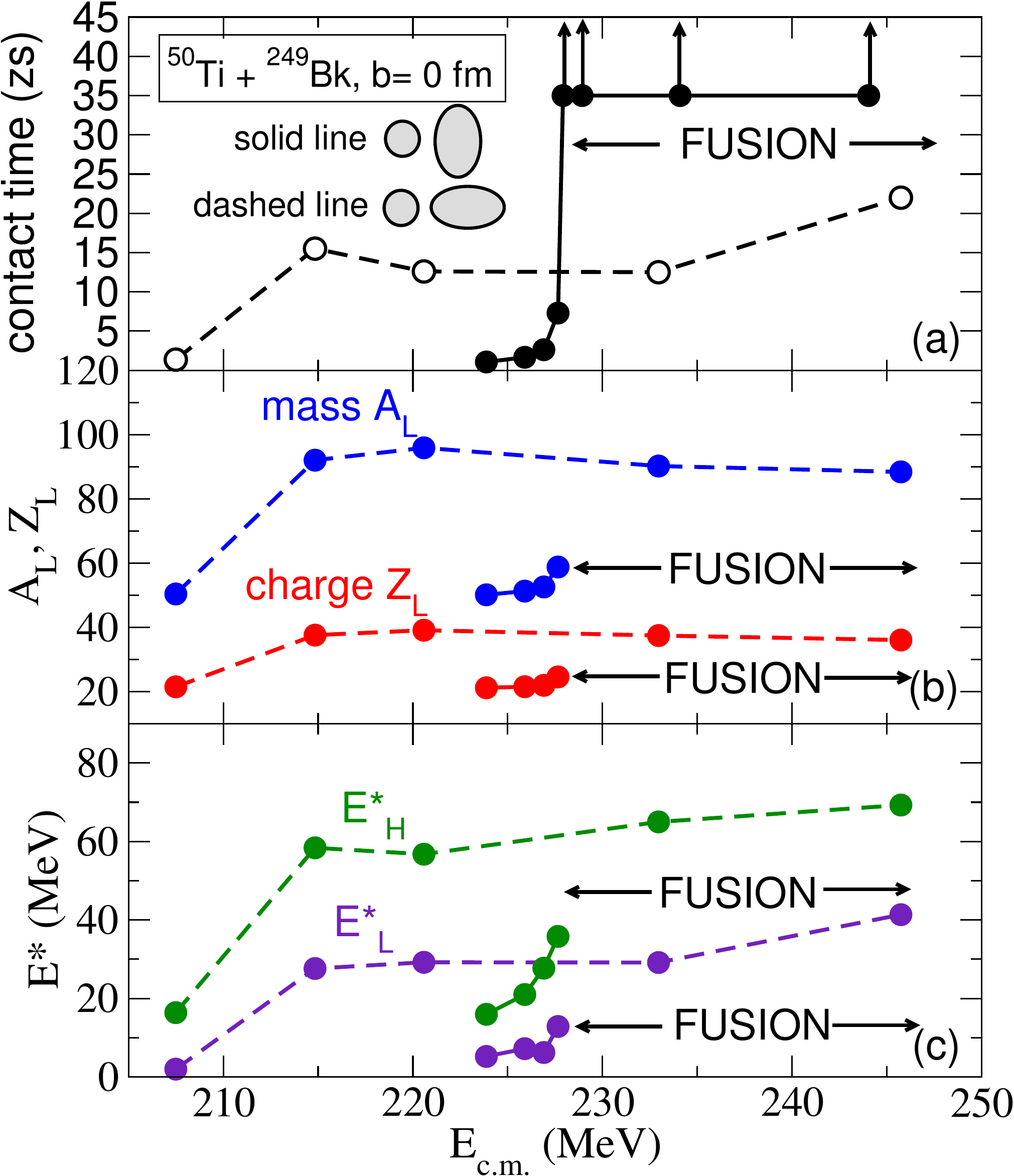}
\caption{\protect\footnotesize (a) Contact time,
(b) mass and charge of the light fragment, and (c) excitation energy $E^{*}$
of the heavy and light fragments
as a function of $E_\mathrm{c.m.}$ for central collisions
of $^{48}$Ca with $^{249}$Bk (left) and $^{50}$Ti with $^{249}$Bk (right). Solid lines are for the side
orientation of the deformed $^{249}$Bk nucleus, and dashed
lines are for the tip orientations.\label{fig:stack1}}
\end{center}
\end{figure}

Figure~\ref{fig:stack1}(a) (left) shows the contact time as a function
of center-of-mass energy for central collisions of $^{48}$Ca with $^{249}$Bk.
For the tip orientation of the $^{249}$Bk nucleus (dashed line) we observe
contact times of order 10--12~zs which are essentially constant over a wide range
of energies, $E_{\mathrm{c.m.}}=$~191--230~MeV. Only at energies below the
potential barrier, $E_B$(tip)$=192.2$~MeV, do the contact times drop off
very rapidly because these events correspond to inelastic scattering
and few-nucleon transfer reactions. A dramatically different picture
emerges for the side orientation of the $^{249}$Bk nucleus (solid line):
At energies above the barrier $E_B$(side)$=205.4$~MeV, the contact times
rise very steeply with energy and reach values up to $22$~zs at
$E_{\mathrm{c.m.}}=210$~MeV. For energies above this value, %TDDFT predicts
fusion is observed which we define by a 
large contact time 
exceeding $35$~zs and 
a mononuclear shape without a neck.

Figure~\ref{fig:stack1}(b) (left) shows the corresponding mass and charge of the light fragment.
We observe that the mass and charge transfer to the light fragment are
roughly proportional to the nuclear contact time. In particular,
for the side orientation of $^{249}$Bk, we find quasielastic collisions at energies below
$E_\mathrm{c.m.}=204$~MeV. Quasifission is limited to a small range of energies
$E_\mathrm{c.m.}=$~209--211~MeV, whereas for energies above $211$~MeV we find
fusion. Naturally, non-central impact
parameters can show quasifission in the range where we see fusion.
The quasifission results are very different for the tip orientation of $^{249}$Bk,
ranging over a much wider energy domain $E_\mathrm{c.m.}=$~191--230~MeV
with a lower maximum mass and charge transfer compared to
the side orientation of $^{249}$Bk.

We have also used %developed an extension to TDDFT theory via the use of a density constraint 
DCTDHF to calculate the excitation energy of {\it each fragment}
directly from the TDHF density evolution.
This gives us new information on the repartition of the excitation energy between
the heavy and light fragments
which is not available in standard TDHF calculations.
In Fig.~\ref{fig:stack1} (c) we show the excitation energies of the
heavy and light fragments which contain approximately 55--60~MeV
and 30--45~MeV of excitation energy (side
orientation) and 50~MeV and 30~MeV (tip orientation),
respectively, for c.m. energies corresponding to quasifission.

Right panel of Fig.~\ref{fig:stack1} shows the corresponding results for central collisions of
 $^{50}$Ti with $^{249}$Bk.
The contact times and the masses and charges of the light fragment show
a similar behavior as a function of energy as compared
to the $^{48}$Ca $+^{249}$Bk reaction.
For the tip orientation, we find quasifission for
$E_\mathrm{c.m.}\ge214$~MeV, with
excitation energies of $E^{*}_H=$~57--69~MeV for the
heavy fragment and $E^{*}_L=$~27--41~MeV for the light fragment, respectively.
The mass and charge of the fragments indicate a strong influence
of the shell effects in the $^{208}$Pb region, as in reactions with $^{48}$Ca.
However, $N=50$ does not seem to play a role here.
For the side orientation, we find inelastic and multi-nucleon transfer reactions
at energies $E_\mathrm{c.m.}=$~223--227~MeV. Quasifission is confined to an
extremely narrow energy window around $E_\mathrm{c.m.}=$~227.4--227.7~MeV, with
excitation energies of $E^{*}_H\simeq36$~MeV and $E^{*}_L\simeq13$~MeV. At energies
$E_\mathrm{c.m.}>228$~MeV, fusion sets in.

\section{Shape evolution and collective dynamics of quasifission}
The proper characterization of fusion-fission and quasifission is one of the most important tasks in analyzing reactions
leading to superheavy elements.
Experimental analysis of fusion-fission and quasifission fragment angular distributions $W(\theta)$  is commonly expressed in terms of
a two-component expression\,\cite{back1985},
\begin{equation}
W(\theta) = \sum_{J=0}^{J_\mathrm{CN}}\mathcal{F}_J^{(FF)}\left(\theta,K_0(FF)\right) + \sum_{J=J_\mathrm{CN}}^{J_\mathrm{max}}\mathcal{F}_J^{(QF)}(\theta,K_0(QF))\;.
\label{eq:wtheta}
\end{equation}
Here, $J_\mathrm{CN}$ defines the boundary between fusion-fission and quasifission,
assuming a sharp cutoff between the angular momentum distributions of each mechanism.
The quantum number $K$ is known to play an important role in fission.
It is the projection of the total angular momentum onto the deformation axis.
In the Transition State Model (TSM),
the characteristics of the fission fragments are determined by the $K$ distribution at scission.
The argument $K_{0}$ entering Eq.~(\ref{eq:wtheta}) is the width of this distribution which is assumed to be Gaussian.
It obeys
$
K_{0}^{2}=T\Im _{eff}/\hbar ^{2}\;,
$
where the effective moment of inertia, $\Im _{eff}$, is computed from the
moments of inertia for rotations around the axis parallel and perpendicular to the %nuclear symmetry axis
principal deformation axis
%\begin{equation}
$\frac{1}{\Im _{eff}}=\frac{1}{\Im _{\parallel }}-\frac{1}{\Im _{\perp }}\;,$
%\label{eq:ieff}
%\end{equation}
and $T$ is the nuclear temperature at the saddle point.
The physical parameters of the fusion-fission part are relatively well known
from the liquid-drop model\,\cite{sierk1986}.
In contrast, the quasifission process never reaches statistical equilibrium.
In principle, it has to be treated dynamically, while Eq.~(\ref{eq:wtheta}) is based on a statistical approximation.
In addition, the usual choice for the nuclear moment of inertia for
the quasifission component, $\Im _{0}/\Im _{eff}=\text{1.5}$\,\cite{yanez2013}, is
somewhat arbitrary. Here, $\Im _{0}$ is the moment of
inertia of an equivalent spherical nucleus.

We have developed methods to extract the moment of inertia of the system (the main collective observable of interest for fission and quasifission)
 directly from TDHF 
time-evolution of collisions resulting in quasifission.
The proper way to calculate the moment-of-inertia for such
time-dependent densities (particularly for non-zero impact parameters)
is to directly diagonalize the moment-of-inertia
tensor represented by a $3\times 3$ matrix with elements
\begin{equation}
\Im_{ij}(t)/m = \int~d^3r\;\rho(\mathbf{r},t) (r^2\delta_{ij}-x_ix_j)\;,
\end{equation}
where $\rho$ is the local number-density calculated from TDHF evolution, %in units of $(fm^{-3})$,
$m$ is the nucleon mass, and $x_{i=1,2,3}$ denote the Cartesian coordinates.
Numerical diagonalization the matrix $\Im$ gives three eigenvalues.
One eigenvalue corresponds to the moment-of-inertia $\Im_{\parallel}$ for the nuclear system rotating
about the principal axis. The other two eigenvalues define the moments of inertia for
rotations about axes perpendicular to the principal axis.
Using the time-dependent moment-of-inertia obtained from the TDHF collision
one can calculate the so-called effective moment-of-inertia defined above.
We have calculated the moment-of-inertia ratio for the $^{48}$Ca~+~$^{249}$Bk non-central collisions
at $E_\mathrm{c.m.}=218$~MeV.
At the point of final touching configuration the moment-of-inertia ratios are in the range 1.4-1.8, suggesting a
relatively strong impact parameter dependence 
which should be accounted for in future extensions to the TSM.

\section*{Acknowledgments}
This work has been supported by the U.S. Department of Energy under grant No.
DE-SC0013847 with Vanderbilt University and by the
Australian Research Councils Future Fellowship (project number FT120100760) and Discovery Projects (project number DP160101254) funding schemes.

\bibliographystyle{ws-procs9x6} % for numbered citation & references
\bibliography{Sanibel_2016.bib}
\end{document}